\documentclass[proof]{pasj00}
\newif\ifAMStwofonts

\def\pmb#1{\mbox{\boldmath$#1$}}
\def\gtsim {>\kern-1.2em\lower1.1ex\hbox{$\sim$}}
\def\ltsim {<\kern-1.2em\lower1.1ex\hbox{$\sim$}}
\def\gtsim {>\kern-1.2em\lower1.1ex\hbox{$\sim$}}
\def\ltsim {<\kern-1.2em\lower1.1ex\hbox{$\sim$}}
\def\ref{\hangindent=1pc \hangafter=1 \noindent}

\def\be{\begin{equation}}
\def\ee{\end{equation}}

\usepackage{epsfig}
\begin{document}

\title{Viscous decretion discs around rapidly rotating stars}

\author{Umin \textsc{Lee}} 
\affil{Astronomical Institute, Tohoku University, Sendai, Miyagi 980-8578, Japan}
\email{lee@astr.tohoku.ac.jp}

\date{Typeset \today ; Received / Accepted}
\maketitle

\begin{abstract}
We discuss steady viscous Keplerian decretion discs around rapidly rotating stars.
We assume that low frequency modes, which may be 
excited by the opacity bump mechanism, convective motion in the core, or tidal force if the star is in a binary system, 
can transport an enough amount of angular momentum to the region close to the stellar surface.
Under this assumption,
we construct a star-disc system, in which there forms a 
viscous decretion disc around a rapidly rotating star because of the angular momentum supply.
We find a series of solutions of steady viscous decretion discs around a rapidly rotating star
that extend to $R_{\rm disc}\gtsim~10R_*$ with $R_*$ being the equatorial radius of the star, depending on
the amount of angular momentum supply.

\end{abstract}

\section{Introduction}

Discs around Be stars are now believed to be viscous Keplerian decretion discs (e.g., Porter \& Rivinius 2003;
Lee, Saio, \& Osaki 1991), although formation mechanism for the decretion discs have not yet been identified.
There have appeared, however, several promising scenarios for the formation mechanism.
For example, stellar evolution calculations of rapidly rotating main sequence stars (e.g., Ekstr\"om et al 2008;
Granada et al 2013) have suggested that
the rotation velocity of the surface layers can reach the critical velocity for mass shedding from the equatorial regions
as a result of angular momentum transfer from the inner region to the surface, where the transport of angular momentum in the evolution calculation
is implemented as in Meynet \& Maeder (2005) employing the theory of the transport mechanisms
in rotating stars developed by Zahn (1992) and Maeder \& Zahn (1998).
Cranmer (2009), on the other hand, has proposed 
a mechanism of angular momentum transfer by waves propagating and damping in the evanescent atmosphere, assuming that
the waves are driven by oscillation modes below the photosphere.
Cranmer (2009) has shown that as a result of angular momentum deposition by waves, the rotation velocity in the atmospheric layers is accelerated to
the Keplerian velocity, leading to the formation of a decretion disc around the rotating star.

Angular momentum transfer by non-axisymmetric oscillations in rotating stars has been discussed by various authors.
Circularization of binary orbit and 
synchronization between the spin of a massive star and the orbital motion of the companion in a binary system
are believed to result from angular momentum exchange between the orbital motion and the spin of the star, where
angular momentum redistribution in the star takes place through dissipative low frequency $g$-modes
tidally excited by the orbital motion of the companion 
(e.g., Zahn 1975, 1977; Goldreich \& Nicholson 1989; Papaloizou \& Savonije 1997; Witte \& Savonije 2001; Willems, van Hoolst, \& Smeyers 2003).
The problem of angular momentum transport in the Sun by low frequency $g$-modes has long been a hot topic in the field 
since the early years of helio-seismology and has been
investigated by many researchers including 
Press (1981), Schatzman (1993), Gough (1997), Kumar, Talon, \& Zhan (1999), Talon, Kumar, \& Zahn (2002), Mathis et al. (2008),
where low frequency modes are assumed to be excited by convection motion 
in the convective envelope and suffer radiative damping as they propagate into the radiative core.
For Be stars, it has also been 
suggested that angular momentum transferred by non-axisymmetric oscillations could
play a role in the formation of viscous discs around the stars
(e.g., Ando 1983, 1986; Lee \& Saio 1993).

As a model of decretion discs around Be stars,
Okazaki (2001) has calculated steady and transonic viscous flows around the stars, taking account of the effects
of the radiation pressure on the radial flow (see also a more recent discussion by Krti\u cka, Owocki, Meynet 2011).
He has shown that the velocity field in the disc is very close to the Keplerian near the star and tends to
angular momentum conserving in the region far from the star.
In his numerical analysis, the disc was treated as being mechanically decoupled of the central star 
and no angular momentum source to support the disc was specified.
These two points are what we are concerned with in this paper.

We construct a disc-star model, consisting of a rapidly rotating star and a viscous decretion disc around it.
To support a decretion disc around a rotating star an enough amount of angular momentum must be supplied 
to the surface layers of the star, and 
this angular momentum supply is assumed to result from angular momentum deposition through
non-axisymmetric oscillation modes.
The rotation velocity in the surface layers is accelerated acquiring angular momentum 
transferred by non-axisymmetric low frequency modes and 
a viscous Keplerian decretion disc forms around the star with the excessive angular momentum.
Since the $\kappa$-mechanism associated with the iron opacity bump
excites low frequency and high radial order $g$-modes and $r$-modes
in slowly pulsating B (SPB) stars and
low radial order $f$- and $p$-modes in $\beta$ Cephei stars (Dziembowski et al. 1993, Gautschy \& Saio 1993),
we may use the low frequency modes as an agent which causes angular momentum transfer in SPBe stars.
For early Be stars, for which the opacity bump mechanism does not work for driving low frequency modes,
we could use low frequency modes stochastically excited by convective motion in the core
(Neiner et al 2012).
If the Be star is in a binary system, low frequency modes tidally excited by the orbital motion of the companion
star could work for disc formation.

We employ a theory of wave-meanflow interaction to derive a meanflow equation for rotation, which 
describes the angular momentum transfer by waves
(e.g., Andrews \& McIntyre 1978ab; Dunkerton 1980; Grimshaw 1984).
We regard the forcing term in the meanflow equation as a source term for angular momentum.
In the angular momentum conservation equation for a disc-star system, therefore, we include
both the forcing term in the wave-meanflow equation and
the viscous torque term, which is essential for angular momentum transfer in the disc.
In \S 2, we derive a set of ordinary differential equations we solve for a steady disc-star system.
\S 3 is for numerical results obtained for steady disc-star systems, and 
we conclude in \S 4.
In the Appendix we derive the meanflow equation we use in this paper.

\section{Equations for Viscous Decretion Discs around Rotating Stars}

To discuss angular momentum transfer by non-axisymmetric oscillations in a rotating star, we use
a theory of wave-meanflow interaction, in which fluid motions are separated into waves and the meanflow and
dissipative processes of the waves propagating in the mean flow
have an essential role for the forcing on the meanflow 
(e.g., Andrews \& McIntyre 1978ab; Dunkerton 1980; Grimshaw 1984; Ando 1983; Goldreich \& Nicholson 1989).
For rotating stars, we regard the rotation velocity field as the meanflow and global oscillations as waves.
In the Cowling approximation (Cowling 1941), in which the Euler perturbation of the gravitational potential is neglected,
the meanflow equation may be given by (see the Appendix)
\begin{eqnarray}
\bar\rho{d\left<\hat\ell\right>\over dt}={m\over 2}{\rm Im}\left[\nabla\cdot\left(\pmb{\xi}^*p^\prime\right)\right],
\end{eqnarray}
where $\pmb{\xi}^*$ is the complex conjugate of the displacement vector $\pmb{\xi}$ of an oscillation mode,
$p^\prime$ is the Eulerian pressure perturbation, $m$ is the azimuthal wave number,
$\bar\rho$ is the mass density in the equilibrium state used for linear treatment of stellar pulsations, and
$\hat\ell$ is the specific angular momentum of rotation.
In the wave-meanflow interaction theory,
the Eulerian coordinates $\hat x_\alpha$ are separated into two parts such that 
$\hat x_\alpha=x_\alpha+\xi_\alpha(t,\pmb{x})$, 
where $x_\alpha$ are the Lagrangian mean coordinates in the sense that
$\left<\xi_\alpha\right>=0$ with $\xi_\alpha$ being the displacement vector component associated with the wave, 
and $\left<f\right>\equiv(2\pi)^{-1}\int_0^{2\pi}f d\phi$ is the zonal average of a physical quantity $f$
with $\phi$ being the azimuthal angle around the rotation axis (see the Appendix).
The right-hand-side of equation (1) represents the forcing effects caused by the oscillation mode,
and an example of the forcing term calculated for low frequency modes in a slowly pulsating B star is
given in the Appendix.

To treat both a rotating star and a geometrically thin viscous disc around it as one system, 
we work in cylindrical coordinates $(R,\varphi,z)$, where
we have omitted the hat $~\hat{}~$ from the Eulerian coordinates $\hat x_\alpha$ for simplicity.
The displacement vector of an oscillation mode may be given in cylindrical coordinates
by $\pmb{\xi}=\xi_R\pmb{e}_R+\xi_\varphi\pmb{e}_\varphi+\xi_z\pmb{e}_z$, where $\pmb{e}_R$, $\pmb{e}_\varphi$, and $\pmb{e}_z$
are orthonormal vectors in the $R$, $\varphi$, and $z$ directions, respectively.
If we regard equation (1) as an angular momentum conservation equation,
its right-hand-side can be considered as a source term for angular momentum attributable to the oscillations.
Including both the viscous torque term for decretion discs and the angular momentum source term due to the oscillations, 
we may write the angular momentum conservation equation as
\be
\rho{d\over dt}\left(R v_\varphi\right)={1\over R}{\partial\over\partial R}\left(R^2\sigma_{R\varphi}\right)
+{m\over 2}{\rm Im}\left[{1\over R}{\partial\over\partial R}\left(R\xi_R^*p^\prime\right)
+{\partial\left(\xi_z^*p^\prime\right)\over \partial z}\right],
\ee
where $d/dt$ denotes the substantial time derivative, and
$\rho$ is the mass density, $v_\varphi$ is the azimuthal velocity,
$\sigma_{R\varphi}$ is the $R\varphi$ component of the viscous tensor and is assumed to dominate 
other components of the tensor for a geometrically thin disc, and we have assumed the system is axisymmetric.

We assume that the oscillation amplitudes are saturated, 
for example, by non-linear couplings between the oscillation modes (e.g., Lee 2012) so that the source term in equation (2) be time-independent.
Assuming further that the disc-star system is in a steady state, and integrating vertically equation (2), we obtain
\begin{eqnarray}
\dot M{\partial\over\partial R}\left(R^2\Omega\right)={\partial\over\partial R}\left(2\pi R^3{\partial\Omega\over\partial R}
\int_{-z_0}^{z_0}\eta dz\right)
+\pi  m{\partial\over\partial R}
R\int_{-z_0}^{z_0} dz~{\rm Im}\left(\xi_R^*p^\prime
\right),
\label{eq:8}
\end{eqnarray}
where we have used $\sigma_{R\varphi}=\eta R(\partial\Omega/\partial R)$ with $\eta$ being the shear viscosity coefficient,
$v_\varphi=R\Omega$, 
and 
we have assumed $p^\prime=0$ at the surface given by $z=z_0(R)$ (see below).
Here, $\dot M\equiv 2\pi R\Sigma v_R$ is the mass decretion rate, which is a constant for steady flows,
and $\Sigma=\int_{-z_0}^{z_0}\rho dz$ is the surface density.
Integrating equation (3) with respect to the coordinate $R$, we obtain
\be
q(R)\equiv 2\pi R^3{\partial\Omega\over\partial R}\int_{-z_0}^{z_0}\eta dz=\dot Mj(R)-f(R)-\dot J,
\label{eq:9}
\ee
where
\be
j(R)=R^2\Omega, 
\ee
\be
f(R)=m\pi  R\int_{-z_0}^{z_0} dz~{\rm Im}\left(\xi_R^*p^\prime
\right),
\ee
and
\be
\dot J\equiv\dot Mj_0-q_0-f_0=\dot Mj-q-f,
\ee
and $j_0$, $q_0$, and $f_0$ denote the quantities evaluated at some arbitrary point $R_0$, which could be, for example, $R_{\rm tr}$
(see below for the definition of $R_{\rm tr}$).

Following Paczy\'nsky (1991), we assume the disc-star system is in hydrostatic balance:
\be
{1\over \rho}{\partial p\over \partial R}+{\partial\Phi\over\partial R}=R\Omega^2,
\ee
\be
{1\over\rho}{\partial p\over\partial z}+{\partial\Phi\over\partial z}=0,
\ee
where the gravitational potential is given by
$
\Phi=-{GM_*/\sqrt{R^2+z^2}}
$
with $M_*$ being the mass of the star.
To determine the surface shape of the disc-star system, we
consider two neighboring points $(R,z_0)$ and $(R+\delta R,z_0+\delta z_0)$ on the surface such that
$p(R,z_0)=0=p(R+\delta R,z_0+\delta z_0)$, and we obtain, using equations (8) and (9) for hydrostatic balance, 
\be
{dz_0\over dR}={R\over z_0}\left({r_0^3\Omega^2\over GM_*}-1\right),
\label{eq:13}
\ee
where $r_0\equiv\sqrt{R^2+z_0^2}$.

Equations (4) and (10) are two ordinary differential equations we solve with appropriate boundary conditions
for disc-star systems.
To treat a disc-star system, we divide the system into
two parts, that is, the inner part ($R\le R_{\rm tr}$) and the outer part ($R\ge R_{\rm tr}$).
The inner part is assumed to be uniformly rotating at a constant rate $\Omega_s$ and 
we have no need to integrate equation (4) for the inner part.
The surface shape of the inner part is obtained by integrating equation (10) for the constant rate $\Omega=\Omega_s$.
On the other hand, the outer part of the system is composed of the outer part of the rotating star and
a decretion disc and is allowed to rotate differentially.
We assume the outer disc part extends to the radius $R_{\rm out}\gg R_{\rm cr}$. 
We therefore have to integrate both
equations (4) and (10) to determine the rotation rate $\Omega(R)$ and the shape $z_0(R)$.

To determine the inner part of the system, let us consider a star uniformly rotating at a rate $\Omega_s$
with no decretion discs around it.
Rewriting equation (10) as
$
{dr_0^2/ dR^2}=r_0^3{\Omega_s^2/ GM_*},
$
we may integrate this equation, since $\Omega_s^2/GM_*$ is a constant for uniform rotation,
to obtain
\be
z_0^s(R)\equiv \left[\left({1\over R_p}-{\Omega_s^2R^2\over 2GM_*}\right)^{-2}-R^2\right]^{1/2},
\ee
where $z_0^s(R)$ defines the surface of the uniformly rotating star (without decretion discs), 
and $R_p=z_0^s(0)$ is its polar radius.
Using the condition $z_0^s(R_e)=0$, we may define, as a function of $R_p$ and $\Omega_s$, 
the equatorial radius $R_e$ of a star that is uniformly rotating at a rate $\Omega_s$.
With the condition $z_0^s(R_{\rm cr})=0$, we can also define the critical equatorial radius $R_{\rm cr}$ of a star uniformly rotating 
at the critical angular velocity $\Omega_{\rm cr}\equiv (GM_*/R_{\rm cr}^3)^{1/2}$.
For these critical radius $R_{\rm cr}$ and rotation rate $\Omega_{\rm cr}$, we have $R_{\rm cr}/R_p=1.5$.
Using this critical radius, we rewrite (11) as
\be
y_0(x)=\left[\left({1\over y_p}-{\bar\Omega_s^2x^2\over 2}\right)^{-2}-x^2\right]^{1/2},
\ee
where $x=R/R_{\rm cr}$, $y_0(x)={z_0^s(R)/ R_{\rm cr}}$, $y_p=R_p/R_{\rm cr}=2/3$, and $\bar\Omega_s=\Omega_s/\Omega_{\rm cr}$.
Note that $y_0(x)=0$ gives the critical radius $x=1$ at $\bar\Omega_s=1$ for a uniformly rotating star.
The rotation rate $\Omega_s$ may also be normalized by using $\Omega_e\equiv (GM_*/R_e^3)^{1/2}$, which is
a critical angular velocity for the actual stellar equatorial radius $R_e$, and in this case
we have $\Omega_s/\Omega_e=x_e^{3/2}\bar\Omega_s$ with $x_e=R_e/R_{\rm cr}$, where $x_e$ is determined as the solution
to $y_0(x)=0$ for a given $\bar\Omega_s\le1$.
Figure 1 plots $x_e$, $\Omega_s/\Omega_e$ and the velocity ratio $V_e/V_{\rm cr}$ as a function of $\bar\Omega_s$,
where $V_e=R_e\Omega_s$ and $V_{\rm cr}=R_{\rm cr}\Omega_{\rm cr}$.
As $\bar\Omega_s$ decreases, $x_e$ tends to 2/3. 
We also note the rapid decrease of $\Omega_s/\Omega_e$ and $V_e/V_{\rm cr}$ as $\bar\Omega_s$ decreases from 1.

\begin{figure}
\resizebox{0.5\columnwidth}{!}{
\includegraphics{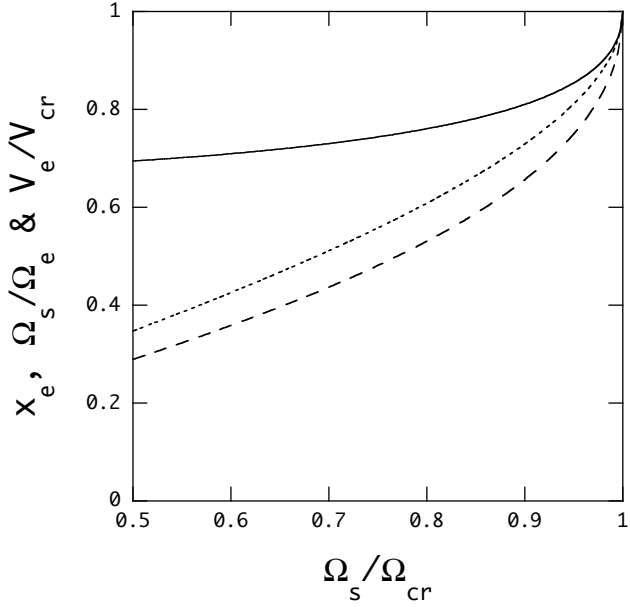}}
\caption{$x_e$ (solid line), $\Omega_s/\Omega_e$ (dashed line) and $V_e/V_{\rm cr}$ (dotted line) as a function of $\bar\Omega_s=\Omega_s/\Omega_{\rm cr}$ for a star uniformly rotating at $\Omega_s$, where
$\Omega_e=(GM_*/R_e^3)^{3/2}$, $\Omega_{\rm cr}=(GM_*/R_{\rm cr}^3)^{3/2}$, $V_e=R_e\Omega_s$, $V_{\rm cr}=R_{\rm cr}\Omega_{\rm cr}$,
and $R_e$ and $R_{\rm cr}$ are the equatorial radius of
a star uniformly rotating at $\Omega_s$ and $\Omega_{\rm cr}$, respectively, and $M_*$ denotes the mass of the star.}
\end{figure}

To determine the outer part of the system by integrating equations (4) and (10), we have to give a prescription
for the viscous angular momentum transport.
For a thin disc in $R\ge R_{\rm tr}$, employing the so called $\alpha$-prescription,
we may give the shear viscosity coefficient $\eta$ as (Shakura \& Sunyaev 1973;
see also Frank, King, \& Raine 2002)
\be
\eta=\alpha \rho_0 v_{s,0}z_0, 
\ee
where $\alpha$ is a dimensionless constant parameter such that $0<\alpha<1$, and $\rho_0$ and 
$v_{s,0}=(\partial p_0/\partial\rho_0)^{1/2}$ are the density and the sound speed evaluated at the equatorial plane.
We note that there exist other ways of prescribing the $\alpha$-viscosity, for example, the $R\varphi$ component of the viscous tensor
is given by $\sigma_{R\varphi}=-\alpha\int_{-z_0}^{z_0}pdz$, which prescription was employed to calculate transonic viscous decretion flows, for example, by Okazaki (2001).
Although this prescription has an advantage that the rank of differential equations can be reduced by one, we use the 
prescription (13) to calculate expected steep changes in $\Omega$ in the boundary layers between
the star and disc.
Assuming a polytropic relation $p=K\rho^{{1+1/ n}}$ and $z_0\ll R$, we
integrate equation (9) to obtain (Paczy\'nsky 1991)
\be
\rho=\rho_0\left(1-{z^2\over z_0^2}\right)^n \quad {\rm with} \quad \rho_0=\left[{GM_*\over 2(n+1)KR^3}\right]^nz_0^{2n},
\ee
and hence
\be
v_{s,0}=\left({GM_*\over 2n}\right)^{1/2}{z_0\over R^{1.5}},
\ee
and
\be
\eta=\alpha c_0{(GM_*)^{n+0.5}\over K^n}{z_0^{2n+2}\over R^{3n+1.5}},
\ee
where $c_0=(2n)^{-1/2}(2n+2)^{-n}$.
Equation (4) is now given by
\be
4\pi\alpha c_0 {(GM_*)^{n+0.5}\over K^n}{z_0^{2n+3}\over R^{3n-1.5}}{d\Omega\over dR}=\dot M R^2\Omega-f-\dot J.
\ee
Equations (10) and (17) now make a set of ordinary differential equations we have to solve with
boundary conditions imposed at $R=R_{\rm tr}$ and $R=R_{\rm out}$.

In this paper, we assume $n=1.5$ for the polytropic index (Paczy\'nsky 1991). 
Using non-dimensional variables, we rewrite equations (10) and (17) as
\be
{dy\over dx}={x\over y}\left[\bar\Omega^2\left(x^2+y^2\right)^{3/2}-1\right],
\ee
\be
{d\bar\Omega\over dx}={1\over a}{x^3\over y^6}\left[x^2\bar\Omega-b \hat f(x)
-\dot j\right],
\ee
where
\be
y={z_0(R)\over R_{\rm cr}}, \quad \bar\Omega={\Omega\over\Omega_{\rm cr}},
\quad a=4\pi\alpha c_0{G^2M_*^2\over \dot MK^{1.5}},
\quad
b={3\over 4}{M_*\Omega_{\rm cr}\over \dot M}, \quad \dot j={\dot J\over \dot M\sqrt{GM_*R_{\rm cr}}},
\ee
and 
\be
\hat f(x)=x\int_{-z_0}^{z_0} {dz\over R_{\rm cr}}{\rho gr\over\bar\rho_* GM_*/R_{\rm cr}}x\sum_km{\rm Im}\left({\xi_R^*\over r}{p^\prime\over \rho gr}
\right),
\ee
and $\bar\rho_*=M_*/(4\pi R_{\rm cr}^3/3)$, $g=GM_r/r^2$ with $M_r$ being the mass within the sphere of radius $r\equiv\sqrt{R^2+z^2}$, 
and the summation in equation (21) is over the oscillation modes
which contribute to the forcing on the meanflow, where $k$ denotes a collective mode index.
For the polytropic index $n=1.5$, we use $K=0.4242GM_*^{1/3}R_*$ with $R_*$ being the radius of the star (Chandrasekhar 1939) to
obtain
\be
a\simeq4\pi\alpha c_0^\prime\left({M_*\over \dot M}\right)\left({GM_*\over R_{\rm cr}^3}\right)^{1/2},
\ee
and, from equation (14),
\be
\rho_0\sim1.5\bar\rho_* ~y^3/x^{9/2},
\ee
where  $c_0^\prime=c_0/(0.4242)^{1.5}\simeq 0.2$ for $n=1.5$ and 
we have replaced $R_*$ by $R_{\rm cr}$, which may be in between $R_*$ and $\sim 1.5R_*$ for rapidly rotating stars.
Although the constant $a$ depends on the viscosity parameter $\alpha$, the magnitude of $a/c_0^\prime$ is almost the same
as that of the constant $b$, that is, $a/c_0^\prime\sim b$.
As suggested by Paczy\'nsky (1991), the constant $a$ can be as large as $a\sim 10^{12}\alpha$, depending 
on the quantities such as $\dot M$, $M_*$, and $R_*$ (or $R_{\rm cr}$).
As $a$ increases, however, it becomes difficult to numerically find solutions to the set of differential equations.
In this paper, we employ $a=10^7$, which leads to the disc thickness $z_0/R\sim 0.1$ (see below).

For rapidly rotating SPB stars, for example, numerous  $r$-modes and prograde sectoral $g$-modes are destabilized by the opacity bump mechanism (e.g., Aprilia, Lee, \& Saio 2012).
To determine the forcing function $b\hat f(x)$ for the SPB stars, we need to know
their amplitudes and to sum up
all the accelerating and decelerating contributions to the forcing.
With a linear theory of oscillations, however, we have no means to determine the amplitudes and hence 
the forcing function $b\hat f(x)$.

In this paper, therefore, we just assume a simple form for the function $b\hat f(x)$.
Assuming that the contributions to acceleration of the surface layers are dominant over those to
deceleration, 
we employ for the forcing function $\hat f(x)$ a form 
given by
\be
b\hat f(x)=b_0\left({1\over \exp\left[-b_1(x-x_0x_e)\right]+1}-1\right)\equiv b_0\bar f(x),
\ee
where $b_0$, $b_1$, and $x_0$ are parameters, and $x_e=R_e/R_{\rm cr}$.
The parameter $b_0$ corresponds to the square of the oscillation amplitudes, and 
we have chosen the functional form for $\bar f(x)$ so that $d\bar f(x)/dx$ roughly reproduces
the $x$ dependence of $1/\tau^{\rm AM}$ shown in Figure 6.
In the surface region of the star where $x\simeq x_e\le 1$, we may approximate 
$|\hat f(x_1\equiv x_0x_e)|\sim x_1^2y_0(x_1)(\rho_0/\bar\rho_*)\left|\sum_km{\rm Im}[(\xi^*_R/r)(p^\prime/\rho gr)]\right|\sim
x_1^{-2.5}y_0^4(x_1)|\sum_kmf_k^2\sin\delta_k|$ where $f_k\sim|\xi_R/R_*|$ is the normalized oscillation amplitude and $\delta_k$ is the phase difference between $\xi_R$ and $p^\prime$ near the
surface, and we have approximated $gr/(GM_*/R_{\rm cr})\sim 1$.
Since $|\bar f(x_1)|= 0.5$, we may have $b_0\sim b\times x_1^{-2.5}y_0^4(x_1)|\sum_kmf_k^2\sin\delta_k|$.
Although it is difficult to correctly estimate the magnitudes of the quantities such as $|\sum_kmf_k^2\sin\delta_k|$,
assuming $|\sum_km\sin\delta_k|\sim 1$ and $x_1^{-2.5}y_0^4(x_1)\sim 10^{-3}$, 
we have $b_0\sim 1$ for $f_k\sim 10^{-2}$ and $b\sim a/c_0^\prime \sim 10^8$ for $a=10^7$.
Note that for $x\gg 1$, we should have $b\hat f(x)=0$, expecting there occurs no forcing on the meanflow
in the disc.

The set of differential equations (18) and (19) for the outer part are integrated with three boundary conditions,
one given by $d\bar\Omega/dx=d\bar\Omega_K/dx$ at $x_{\rm out}\equiv R_{\rm out}/R_{\rm cr}$ 
and other two conditions given by $\bar\Omega(x_{\rm tr})=\bar\Omega_s$ and $y(x_{\rm tr})=y_0(x_{\rm tr})$ 
at $x_{\rm tr}\equiv R_{\rm tr}/R_{\rm cr}$, where $\bar\Omega_K=x^{-3/2}$ is the Keplerian angular velocity.
The first two boundary conditions are used to integrate the set of coupled two first order, ordinary differential equations (18) and (19) for a given $\dot j$, which may be regarded as an eigenvalue of the system of differential equations
and is determined by using the third condition.
The third condition $y(x_{\rm tr})=y_0(x_{\rm tr})$ ensures the physical continuity of the inner part and
the outer part of the system at $x_{\rm tr}$.
In this paper, we use $x_{\rm tr}=0.8$.

As indicated by equation (19), $d\bar\Omega/dx$ changes its sign at $x_j$, at which 
$x_j^2\bar\Omega(x_j)-\dot j=0$ if we assume $b\hat f(x)=0$ for $x> 1$, and 
we obtain $x_j\simeq {\dot j}^2$ if we substitute $\bar\Omega_K$ for $\bar\Omega$.
To understand a rough property of the solution in the region $1< x <x_j$ in which $x^2\bar\Omega\ll \dot j$,
we assume functional forms given by $y=c_1 x^s$ and $\bar\Omega=c_2\bar\Omega_K$, where the parameters $c_1$, $c_2$, and $s$
are assumed only weakly dependent on $x$,
and substituting the forms into equations (18) and (19), we have 
$sc_1x^{2s-2}=c_2^2(1+y^2/x^2)^{3/2}-1$ and $1.5c_2x^{-2.5}=\dot j/(ac_1^6x^{6s-3})$, and hence 
setting $2.5=6s-3$, we obtain
\be
s={11\over 12}, \quad c_1\sim\left({{\dot j}\over ac_2}\right)^{1/6}, 
\quad c_2\sim {\left(1+sc_1^2x^{2s-2}\right)^{1/2}\over\left(1+c_1^2x^{2s-2}\right)^{3/4}}
\sim{1\over \left(1+c_1^2x^{-1/6}\right)^{1/4}},
\ee
where we have approximated $(2/3)^{1/6}\sim 1$ for the second equation and we have set the factor $s$, on the left hand of 
$c_1^2$ in the numerator of the third equation, equal to 1.
These relations are consistently satisfied if $c_1\sim 0.1$ and $c_2\sim 1$ for ${\dot j}\sim 10$ and $a\sim 10^7$, which indicates that the decretion discs are geometrically thin for the parameter values.
If the disc extends to large radii beyond $x_j$, the disc flows may tend to angular momentum conserving so that $x^2\bar\Omega-\dot j$ be a constant
(see Okazaki 2001).
If this is the case, the outer boundary condition must be modified and an appropriate treatment of solutions around a point of $x_j$ will be required.

\begin{table}
\begin{center}
\caption{$x_m$, $\dot j$ and $-b_0\bar f(x_m)$ for decretion disc solutions }
\begin{tabular}{@{}ccccc}
\hline
$\Omega_s/\Omega_{\rm cr}$ & $x_e$ & $x_m$ & ${\dot j}$ & $-b_0\bar f(x_m)$ \\
\hline
 &  $b_0=50$ & $x_0=0.95$ & $x_{\rm out}=10$ &\\
\hline
0.9750 & 0.890 &  0.893 &5.211 & 4.255 \\
0.9800 & 0.900 &  0.905 &4.859 & 3.793\\
0.9850 & 0.912 &  0.919 &4.405 & 3.362\\
0.9900 & 0.926 &  0.936 &3.837 & 2.826\\
0.9950 & 0.946 &  0.957 &3.603 & 2.571\\
0.9980 & 0.965 &  0.963 &5.420 & 4.505\\
0.9990 & 0.975 &  0.964 &7.629 & 6.576\\
1.0000 & 1.000 &  0.965 &16.81 & 16.04\\
\hline
  & $b_0=10$ & $x_0=0.95$ &$x_{\rm out}=2$&\\
\hline
0.9935 & 0.940 &  0.945 & 1.688 & 0.691\\
0.9940 & 0.942 &  0.948 & 1.666 & 0.657\\
0.9950 & 0.946 &  0.953 & 1.619 & 0.621\\
0.9960 & 0.952 &  0.960 & 1.578 & 0.584\\
0.9970 & 0.958 &  0.966 & 1.552 & 0.576\\
0.9980 & 0.965 &  0.972 & 1.577 & 0.594\\
0.9990 & 0.975 &  0.976 & 1.761 & 0.767\\
1.0000 & 1.000 &  0.976 & 3.160 & 2.142\\
\hline
  & $b_0=50$ & $x_0=0.99$ &$x_{\rm out}=10$&\\
\hline
0.9500 & 0.854 &  0.856 &19.41 & 18.56\\
0.9600 & 0.867 &  0.870 &18.77 & 17.91\\
0.9700 & 0.881 &  0.886 &17.82 & 16.70\\
0.9800 & 0.900 &  0.907 &16.39 & 15.50\\
0.9900 & 0.926 &  0.936 &15.10 & 13.81\\
0.9950 & 0.946 &  0.952 &16.86 & 15.79\\
0.9980 & 0.965 &  0.962 &21.97 & 20.88\\
0.9990 & 0.975 &  0.966 &25.70 & 24.53\\
1.0000 & 1.000 &  0.973 &36.16 & 35.03\\
\hline
\end{tabular}
\medskip
\end{center}
\end{table}

\section{Numerical Results}

Let us give a brief description of the procedure we employ to obtain solutions to the set of differential equations (18) and (19) for a given $\bar\Omega_s\le 1$.
Since it is difficult to solve the differential equations for the entire region
from $x_{\rm tr}$ to $x_{\rm out}$ by
using a Runge-Kutta method (or a relaxation method), 
we divide the interval $(x_{\rm tr},x_{\rm out})$ into two intervals, that is, $(x_{\rm tr},x_m)$
and $(x_m,x_{\rm out})$ with $x_m\simeq 1$, and for integration
we use an implicit Runge-Kutta method for the former and a Henyey type relaxation method for the latter.
Here, for $x_m$ we choose a point $x_m\simeq 1$ that satisfies $d\bar\Omega/dx=0$.
For a given value of the parameter ${\dot j}$, 
we integrate the differential equations (18) and (19),
from $x=x_{\rm tr}$ with starting values $y(x_{\rm tr})=y_0(x_{\rm tr})$ and $\bar\Omega(x_{\rm tr})=\bar\Omega_s$, to the point $x_m$.
This integration gives $x_m$, $y_m=y(x_m)$, and $\bar\Omega_m=\bar\Omega(x_m)$ as a function of $\dot j$,
or equivalently, $y_m$, $\bar\Omega_m$, and $\dot j$ as a function of $x_m$.
For the interval between $x_m$ and $x_{\rm out}$, we then solve equations (18) and (19) using the relaxation method 
with the initial guesses given by $y=x[({\dot j}x^{-1/2}-1)/1.5a]^{1/6}$ and $\bar\Omega=x^{-3/2}$
to find the value of $x_m$ such that the boundary conditions
$y(x_m)=y_m$ and $\bar\Omega(x_m)=\bar\Omega_m$ at $x=x_m$ and $d\bar\Omega/dx=d\bar\Omega_K/dx$ at $x=x_{\rm out}$
are satisfied.
This procedure gives us a complete solution $y(x)$ and $\bar\Omega(x)$ for the region
from $x_{\rm tr}$ to $x_{\rm out}$, which corresponds to the outer part of a disc-star system.
The inner part of the system is the part of a star uniformly rotating at the rate $\bar\Omega_s$ and its surface shape $y_0(x)$
is given by equation (12).
The inner part and outer part of the system are connected at $x_{\rm tr}$, and the continuous connection is ensured
by the boundary conditions given by $y(x_{\rm tr})=y_0(x_{\rm tr})$ and $\bar\Omega(x_{\rm tr})=\bar\Omega_s$.

Figure 2 shows $y$ and $\bar\Omega$ as a function of $x$ for $\bar\Omega_s=0.98$, 0.99, and 1.00, where
we have assumed $b_0=b_1=50$, $x_{\rm out}=10$, and $x_0=0.95$.
In Table 1, we tabulate several characteristic quantities such as $x_m$, $\dot j$, and $-b_0\bar f(x_m)$  as a function
of $\bar\Omega_s$.
As shown by the left panel of the figure, there appears a sharp dip in $y$ at the boundary between the star and disc,
and the dip becomes deeper for smaller values of $\bar\Omega_s$.
Table 1 indicates that this star-disc boundary is located at a radius near $x_e$.
If we go outwards from $x_{\rm tr}$, 
$\bar\Omega$ starts at a point near $x_e$ to steeply increase
to attain a super-Keplerian rate ($\Omega>\Omega_K$) at $x_m$ and 
then decreases to the Keplerian velocity.
Note that $\Omega$ is slightly sub-Keplerian in the region of $x\gtsim 1$.
For a given $b_0$, there exists the lower limit of $\Omega_s$, below which no solutions to the differential equations are found.
As $\bar\Omega_s$ decreases from unity, the amount of angular momentum deposition required to accelerate 
the sub-Keplerian rotation velocity to a super-Keplerian one
is increased, and hence the derivative $d\bar\Omega/dx$ inevitably becomes steeper in the region where
the acceleration takes place.
The lower limit of $\bar\Omega_s$ is reached when the point of $y=0$ or $d\bar\Omega/dx=\infty$ appears in the solution $y$ or
$\bar\Omega$.
For $b_0=50$ and $x_{\rm out}=10$, the lower limit of $\bar\Omega_s$ is $\simeq0.975$.

Figure 3 shows that the ratio $y/x$ is less than $\sim 0.1$ in the disc, indicating the disc is geometrically thin.
Since the ratio is approximately proportional to ${\dot j}^{1/6}$ as suggested by equation (25) and
the value of ${\dot j}$ for $\bar\Omega_s=0.98$ is larger than that for $\bar\Omega_s=0.99$ (see Table 1), 
the ratio $y/x$ for the former is larger than that for the latter.
This figure also shows that the ratio gradually decreases as $x$ increases from $x\sim 1$.

If we employ $b_0=10$ instead of $b_0=50$, we can obtain solutions for $x_{\rm out}=2$ but
no solutions for $x_{\rm out}=10$ and the parameter value of $\dot j$ we obtain for $b_0=10$ is $\sim 1.5$ for $\bar\Omega_s <1$.
This suggests that proper solutions to the differential equations can be obtained
only when the outer boundary condition is imposed at $x_{\rm out}\ltsim ~x_j\simeq {\dot j}^2$.
We also find that the properties of the solutions for a given ${\dot j}$ 
do not strongly depend on $x_{\rm out}$ so long as $x_m< x_{\rm out}\ltsim ~x_j$.
(These properties of disc solutions are confirmed also for the case of $b_0=50$.)
For $b_0=10$, the lower limit of $\bar\Omega_s$ is $\simeq 0.9935$, 
which is much closer to unity than the lower limit $\simeq0.975$ for the case of $b_0=50$.
Figure 4 plots $y$ and $\bar\Omega$ as a function of $x$ for $b_0=10$.
The discs for $b_0=10$ are thinner than for $b_0=50$, and the peak value $\bar\Omega(x_m)$ attained for
the lower limit of $\bar\Omega_s$ is smaller than
that for $b_0=50$.

\begin{figure*}
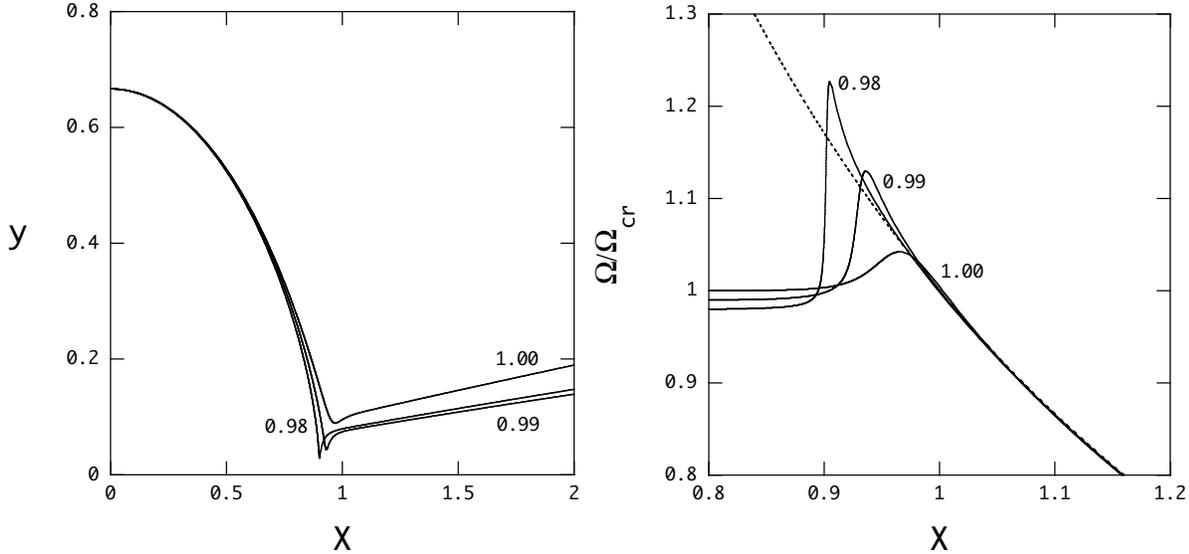

\resizebox{0.45\columnwidth}{!}{
\includegraphics{f2a.epsi}}
\resizebox{0.467\columnwidth}{!}{
\includegraphics{f2b.epsi}}
\caption{$y=z_0(R)/R_{\rm cr}$ (left panel) and $\bar\Omega=\Omega/\Omega_{\rm cr}$ (right panel) as a function of $x=R/R_{\rm cr}$ for
$\bar\Omega_s=0.98$, 0.99, and 1.00 for $b_0=b_1=50$, $x_0=0.95$, and $x_{\rm out}=10$, where
$\Omega_{\rm cr}=(GM_*/R_{\rm cr}^3)^{3/2}$ with $M_*$ being the mass of the star, and the numbers attached to the lines indicate the values of $\bar\Omega_s$.
The dotted line in the right panel indicates the Keplerian rotation rate given by $\bar\Omega_K=x^{-3/2}$.}
\end{figure*}

\begin{figure*}
\resizebox{0.5\columnwidth}{!}{
\includegraphics{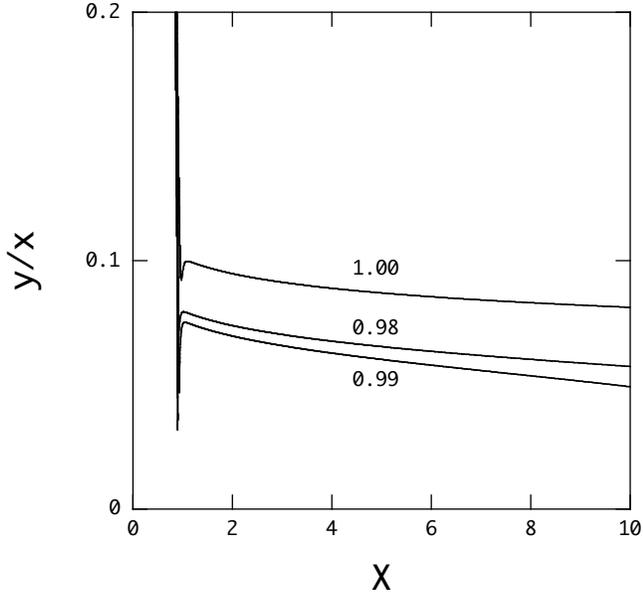}}
\caption{The ratio $y/x$ versus $x=R/R_{\rm cr}$ for $\bar\Omega_s=0.98$, 0.99, and 1.00 for $b_0=b_1=50$ and
$x_0=0.95$, and the numbers attached to the lines indicate the values of $\bar\Omega_s$.}
\end{figure*}

\begin{figure*}
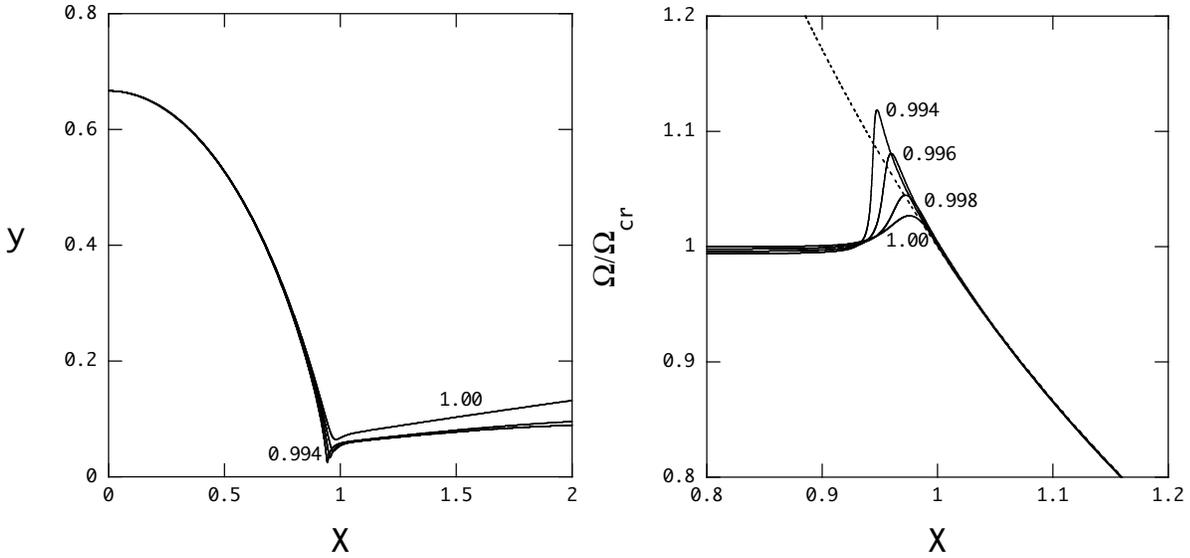

\resizebox{0.45\columnwidth}{!}{
\includegraphics{f4a.epsi}}
\resizebox{0.467\columnwidth}{!}{
\includegraphics{f4b.epsi}}
\caption{$y=z_0(R)/R_{\rm cr}$ (left panel) and $\bar\Omega=\Omega/\Omega_{\rm cr}$ (right panel) as a function of $x=R/R_{\rm cr}$ for
$\bar\Omega_s=0.994$, 0.996, 0.998, and 1.00 for $b_0=10$, $b_1=50$, $x_0=0.95$, and $x_{\rm out}=2$, and the numbers attached to the lines indicate the values of $\bar\Omega_s$.
The dotted line in the right panel indicates the Keplerian rotation rate given by $\bar\Omega_K=x^{-3/2}$.}
\end{figure*}

To examine the case in which the acceleration takes place in a region much closer to the stellar surface, that is, 
in the region of much lower density,
we have carried out similar calculations assuming $x_0=0.99$ for
$b_0=b_1=50$ and $x_{\rm out}=10$.
As shown by Figure 5 and Table 1, 
we again obtain a series of decretion disc solutions, the properties of which are
quite similar to those for $x_0=0.95$, except for that the values of ${\dot j}$ for $x_0=0.99$
are much larger than those for $x_0=0.95$.
Because of the large values of the parameter $\dot j$, the discs can have larger radii for $x_0=0.99$
than for $x_0=0.95$.
The lower limit of $\bar\Omega_s$ is $\simeq 0.95$, which is smaller than $\simeq0.975$ for the case of $x_0=0.95$.

\begin{figure*}
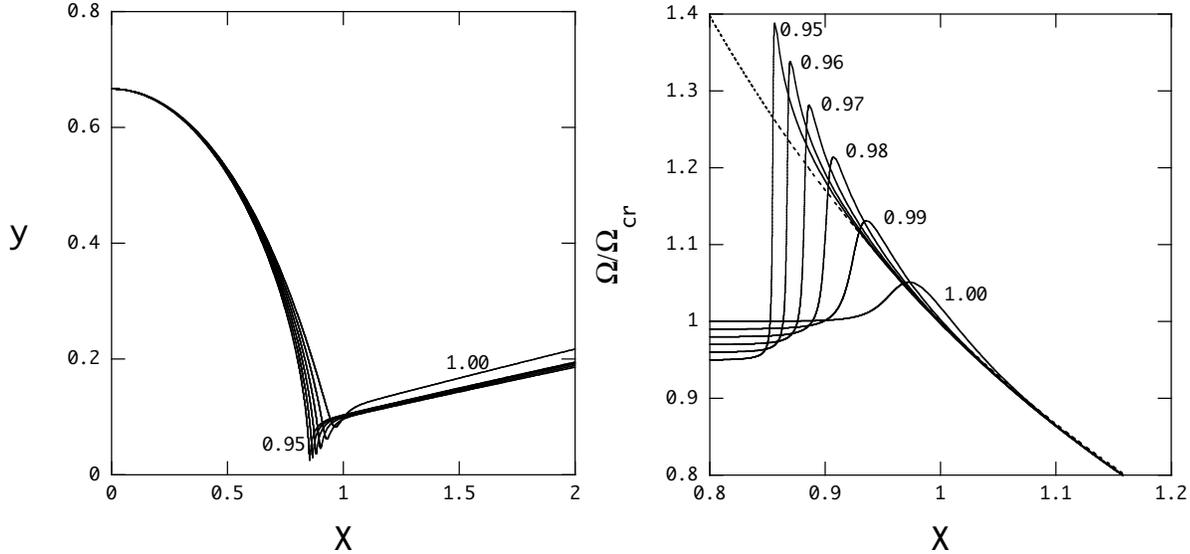

\resizebox{0.45\columnwidth}{!}{
\includegraphics{f5a.epsi}}
\resizebox{0.467\columnwidth}{!}{
\includegraphics{f5b.epsi}}
\caption{$y=z_0(R)/R_{\rm cr}$ (left panel) and $\bar\Omega=\Omega/\Omega_{\rm cr}$ 
(right panel) as a function of $x=R/R_{\rm cr}$ for
$\bar\Omega_s=0.95$, 0.96, 0.97, 0.98, 0.99, and 1.00 for $b_0=b_1=50$, $x_{\rm out}=10$, and $x_0=0.99$, and the numbers attached to the lines indicate the values of $\bar\Omega_s$.
The dotted line in the right panel indicates the Keplerian rotation rate given by $\bar\Omega_K=x^{-3/2}$.}
\end{figure*}

Let us discuss about the physical meaning of $\dot j$.
With the substitution of $R_m$ for $R_0$, equation (7) becomes $\dot J\equiv \dot Mj_m-f_m=\dot Mj(x)-q(x)-f(x)$, where $f_m=f(x_m)$ and $j_m=j(x_m)$.
The quantity $\dot j=x_m^2\bar\Omega(x_m)-b_0\bar f(x_m)$ is now composed of
the angular momentum of rotation
and the excessive angular momentum due to the forcing by the waves at $x_m$.
Since $q(x_{\rm tr})\simeq 0$, we have $\dot M(j_m-j(x_{\rm tr}))\simeq f_m-f(x_{\rm tr})$, which suggests that
the acceleration from $j(x_{\rm tr})$ to $j_m$ is caused by angular momentum deposition
equal to $f_0-f(x_{\rm tr})$.
We also note that the excessive angular momentum $-b_0\bar f(x_m)$ is used to extend the disc outward from $x_m$
where $-b_0\bar f(x)$ tends to zero as $x$ increases from $x_m$.
Since $x_m^2\bar\Omega(x_m)\sim 1$ as suggested by Table 1, the value of $\dot j$ and hence the possible extension of the disc is determined by the amount of this excessive angular momentum $-b_0\bar f(x_m)$.
If $-b_0\bar f(x_m)$ is large, the possible extension of viscous Keplerian decretion discs becomes large.

\section{Conclusion}

We have calculated steady viscous Keplerian decretion discs around a rapidly rotating star, assuming 
the existence of angular momentum supply to the region close to the surface of the star.
The angular momentum supply may be provided by angular momentum deposition that takes place through
wave-meanflow interaction, where the waves are low frequency global oscillations
excited by the opacity bump mechanism for SPB stars, or by a stochastic mechanism for early Be stars, 
or by the tidal force if the star is in a binary system.
We may conclude that the angular momentum supply to the surface layers by the waves 
can be a mechanism for disc formation around rapidly rotating Be stars.
In the sense that angular momentum supply to the surface layers plays an essential role for disc formation, 
our calculation may be thought complementary to recent stellar evolution calculations of rotating main sequence stars by
Granada et al (2013), who suggested that in the course of evolution the surface layers of the rotating stars reach
the critical rotation velocity, leading to mass shedding from the equatorial regions, where
the transport of angular momentum inside a star is implemented following the prescription of Zahn (1992)
for the horizontal diffusion coefficient and that of Maeder (1997) for the 
shear diffusion coefficient.

If the amount of angular momentum supply, which is represented by the parameter $b_0$ in this paper, 
is large enough, viscous decretion discs can extend to a distance
as far as $R_{\rm out}\gtsim ~10R_{\rm cr}$, and if the acceleration takes place in the region very close to the stellar surface, the possible extension 
a decretion disc attain can be as large as $R_{\rm out}\gtsim 100R_{\rm cr}$.
If $b_0$ is small, however, disc solutions are found only when $\bar\Omega_s$ is very close to unity,
and the possible extension of the discs is comparable to the stellar radius $R_{\rm cr}$ itself.

If the angular momentum supply is provided by global oscillations,
the parameter $b_0$ represents the square of the oscillation amplitudes.
We have argued that the amplitudes of order of $\xi_R/R_{*}\sim 0.01-0.1$ can lead to reasonable values of $b_0$.
It is also important to note that for given values of the parameters $\alpha$, $M_*$, $R_*$ (or $R_{\rm cr}$), assigning a value to the parameter $a$ is almost equivalent to assuming a single value for $\dot M$.
The discussions made in this paper, therefore, are those for a single value of $\dot M$, which is determined from the value $a=10^7$ for given $M_*$, $R_*$, and $\alpha$.

The density at the mid-plane of the disc may be estimated by using equation (23), which
leads to $\rho_0\propto x^{-1.5}$ if the ratio $y/x$ is assumed almost constant.
Since $\rho_0\propto x^{-3.5}$ has been suggested observationally, the $x$ dependence of $\rho_0$
in our model is in a serious conflict with the observational estimation (e.g., Porter \& Rivinius 2003).
For B type stars we have a typical mean density $\bar\rho_*\sim 10^{-2}~{\rm g~cm}^{-3}$, 
and since $y/x\sim 0.01 -0.1$, we obtain
$\rho_0\sim 10^{-9} - 10^{-6}~{\rm g~cm}^{-3}$ at $x\sim 10$, the value of which is much higher than that observationally estimated
(e.g., Waters 1986).
We could use much larger (smaller) values for the parameter $a$ ($\dot M$) 
to reduce the ratio $y/x$ and hence $\rho_0$, but for the value of $a$ much larger than $10^7$, 
we find it difficult to numerically obtain solutions to the differential equations.
Note that
decretion discs calculated for $a$ much larger than $10^7$ (i.e., for $\dot M$ much smaller than 
that for $a=10^7$) would have large extensions even for small values of $b_0$, 
although we cannot prove because of the numerical difficulty.

In our steady disc-star systems discussed in this paper, 
the extension of the discs is limited by $x_j\simeq \dot j^2$.
At large radii $x\gg 1$, the disc flows possibly tend to angular momentum conserving ones (e.g., Okazaki 2001),
or the discs would suffer radiative ablation to be truncated at finite radii (e.g., Krti\u cka, Owocki, Meynet 2011).
To obtain steady and angular momentum conserving disc solutions at large radii, 
we need to calculate transonic flows extending indefinitely, and
the set of differential equations we have solved in this paper, however, do not provide such transonic solutions.
We think this is a reason for the differences in the properties,
such as the $x$ dependence of $\rho_0$ discussed in the previous paragraph, 
of viscous disc solutions at large radii between Okazaki (2001) and the present paper.

It is important to note that decretion disc solutions in our model are obtained only for
$\bar\Omega_s$ that is close to 1 (see Table 1), and 
that since $\Omega(x)/\Omega_{\rm cr}\ge1$ in the boundary layers between the disc and star as indicated by Figures 2, 4 and 5,
the actual observed values of $V_{\rm em}/V_{\rm cr}$ will be close to 1 even if $V_e/V_{\rm cr}\sim 0.8-0.9$ (see Figure 1), where we may define $V_{\rm em}= R_m\Omega(x_m)$ with $R_m=x_mR_{\rm cr}$.
Although various attempts (e.g., Townsend, Owocki, Howarth 2004; Cranmer 2005; Fr\' emat et al 2005;
Rivinius, \u Stefl, Baade 2006; Delaa et al 2011) have been made to estimate the ratio $\Omega_s/\Omega_{\rm cr}$ (or $V_e/V_{\rm cr}$)
for Be stars to judge whether Be stars are rotating at rates very close to the critical ones or
at rates substantially lower than the critical rates, it may be fair to say that no firm conclusions
concerning the ratio have been obtained.
For Be stars, for example,
Townsend, Owocki, \& Howarth (2004) argued for the rotation rates very close to the critical rates,
but Fr\' emat et al (2005) estimated the average rate of rotation as $\Omega/\Omega_{\rm cr}\simeq 0.88$,
which may be considered as substantially subcritical rotation rates.
More interestingly, Cranmer (2005) have suggested that the lower limits of the rotation rates
for early type Be stars are as low as 40\%$-$60\% of the critical rates but those for late type
Be stars could be very close to the critical ones.
Since the model discussed in this paper becomes viable only for stars rotating at a rate close to the
critical rate, the model will be ruled out if it is proved that most of Be stars
are rotating at rates much lower than the critical rates.

As indicated by the plots of $\bar\Omega(x)$, there occurs a strong differential rotation in the 
region close to the surface, particularly for lower values of $\bar\Omega_s$.
The strong differential rotation could modify the modal properties of oscillations and hence the accelerating and decelerating contributions to
the forcing on the velocity field.
Stability analysis of low frequency modes in differentially rotating stars, which will be one of our future studies, 
is necessary if we use for the forcing mechanism the oscillation modes that are excited by the opacity bump mechanism.

\begin{appendix}
\section{Mean Flow Equation in the Lagrangian Mean Formalism}

Following Grimshaw (1984), in this Appendix
we derive a meanflow equation for zonal flows around the rotation axis of stars,
using the Lagrangian mean formalism
(see also Andrews \& McIntyre 1978b).
In a frame rotating with the angular velocity $\Omega_c$, the $\phi$ component of
the momentum conservation equation in spherical polar Eulerian coordinates 
$\hat{\pmb{x}}\equiv\left(\hat x_1, \hat x_2, \hat x_3\right)=\left(\hat r,\hat\theta,\hat\phi\right)$ 
is given by
\be
\hat\rho\left[{d\hat v_\phi\over dt}+{\hat v_r \hat v_\phi\over \hat r}+{\hat v_\theta \hat v_\phi\cot\hat\theta\over \hat r}
+2\Omega_c\left(\sin\hat\theta \hat v_r+\cos\hat\theta\hat v_\theta\right) \right]
=-{1\over \hat r\sin\hat\theta}{\partial \hat p\over\partial \hat\phi}-\hat\rho{1\over \hat r\sin\hat\theta}{\partial \hat\Phi\over\partial \hat\phi},
\ee
where $\hat \rho$, $\hat p$ and $\hat \Phi$ are respectively the density, the pressure, and the gravitational potential
of the fluid,
\be
\hat v_r={d\hat r\over dt}, \quad \hat v_\theta=\hat r{d\hat\theta\over dt}, \quad \hat v_\phi=\hat r\sin\hat\theta{d\hat\phi\over dt},
\ee
and
\be
{d\over dt}={\partial\over\partial t}
+\hat v_r{\partial\over\partial \hat r}
+{\hat v_\theta\over \hat r}{\partial\over\partial\hat\theta}
+{\hat v_\phi\over \hat r\sin\hat\theta}{\partial\over\partial\hat\phi}.
\ee
To discuss wave-mean flow interactions, we introduce the Lagrangian mean coordinates 
$\pmb{x}\equiv\left(x_1,x_2,x_3\right)=\left(r,\theta,\phi\right)$ and the displacement vector $\xi_\alpha$
for $\alpha=1,~2,~ 3$ such that
\be
\hat x_\alpha=x_\alpha+\xi_\alpha\left(t,\pmb{x}\right).
\ee
We assume that for any given $\hat v_\alpha$ there is a unique ``reference" velocity $\bar v_\alpha$, such that
when the point $x_\alpha$ moves with velocity $\bar v_\alpha$ the point $\hat x_\alpha$ moves with velocity 
$\hat v_\alpha$ (e.g., Grimshaw 1984).
Using the reference velocity, we may define 
\be
{d\over dt}={\partial\over\partial t}+\bar v_r{\partial\over\partial r}
+{\bar v_\theta\over r}{\partial\over\partial\theta}
+{\bar v_\phi\over r\sin\theta}{\partial\over\partial\phi},
\ee
where $\partial /\partial t$ indicates the partial time derivative with the coordinates $\pmb{x}$ being held constant.
For the wave-mean flow interaction formulatoin for zonal flows, 
we introduce an averaging procedure defined by
\be
\left< f\right>\equiv{1\over 2\pi}\int_0^{2\pi}fd\phi,
\ee
where $\phi$ can be regarded as the ensemble parameter
such that $f(x_\alpha,\phi+2\pi)=f(x_\alpha,\phi)$ (e.g, Grimshaw 1984), and we assume for the displacement $\xi_\alpha$
\be
\left<\xi_\alpha\right>=0.
\ee
The velocity $\bar{\pmb{v}}$ may be regarded as the mean velocity associated with the coordinates $\left(x_\alpha\right)$,
and the displacement $\xi_\alpha$ represents the waves.

Using the Jacobian $J$ for the coordinate transformation between $\left(\hat x_\alpha\right)$ and 
$\left(x_\alpha\right)$ given by
\be
J\equiv\det\left({\partial \hat x_\alpha\over \partial x_\beta}\right)
=\det\left(\delta_{\alpha\beta}+{\partial \xi_\alpha\over \partial x_\beta}\right),
\ee
where $\delta_{\alpha\beta}$ denotes the Kronecker delta,
we define the mean density $\tilde\rho$ associated with the coordinate $\pmb{x}$ as
\be
\hat\rho \hat r^2\sin\hat\theta J=\tilde \rho r^2\sin\theta.
\ee
It is convenient to introduce (e.g., Andrews \& McIntyre 1978b)
\be
K_{\alpha\beta}={\partial J\over \partial \left(\partial \hat x_\alpha/\partial x_\beta\right)}
={1\over 2}\epsilon_{\alpha\kappa\lambda}\epsilon_{\beta\sigma\tau}{\partial \hat x_\kappa\over\partial x_\sigma}
{\partial \hat x_\lambda\over\partial x_\tau},
\ee
for which we have
\be
{\partial K_{\alpha\beta}\over \partial x_\beta}=0,
\ee
and
\be
K_{\alpha\beta}{\partial \hat x_\alpha\over\partial x_\gamma}=\delta_{\beta\gamma}J
=K_{\beta\alpha}{\partial \hat x_\gamma\over\partial x_\alpha},
\ee
where we have employed the dual summation convention that Greek indices are summed over the range 1 to 3, 
and $\epsilon_{123}=\epsilon_{231}=\epsilon_{312}=1=
-\epsilon_{132}=-\epsilon_{213}=-\epsilon_{321}$ and $\epsilon_{\alpha\beta\gamma}=0$ otherwise.

We rewrite equation (A1) as
\be
\hat \rho{d\hat \ell\over dt}=-{\partial\over\partial\hat\phi}\hat p-
\hat\rho{\partial\over\partial\hat\phi}\hat\Phi,
\ee
where $\hat\ell$ is the specific angular momentum, in an inertial frame, around the rotation axis defined by
\be
\hat\ell=\hat r\sin\hat\theta \hat v_\phi+\Omega_c \hat r^2\sin^2\hat\theta.
\ee
Multiplying equation (A13) by $\hat r^2\sin\hat\theta J$, we obtain
\be
\tilde\rho {d\hat\ell\over dt}
+{\partial\tilde p\over\partial\phi}=-{1\over r^2\sin\theta}{\partial\over\partial x_\beta} R_{3\beta}
-\tilde\rho{\partial\over\partial\hat\phi}\hat\Phi,
\ee
where
\be
R_{3\beta}=\delta_{3\beta}\left(J\hat r^2\sin\hat\theta\hat p-r^2\sin\theta\tilde p\right)
-\hat r^2\sin\hat\theta \hat p{\partial\xi_\gamma\over\partial x_3}K_{\gamma\beta},
\ee
and $\tilde p=\hat p\left(\tilde\rho, \hat s\right)$
with $\hat s$ being the specific entropy, and we have used the identity
\be
 K_{3\beta }\hat r^2 \sin \hat\theta \hat p =\delta _{3\beta } J\hat r^2 \sin \hat\theta \hat p 
 -\hat r^2 \sin \hat\theta \hat p\frac{{\partial \xi _\gamma  }}{{\partial x_3 }}K_{\gamma \beta } .
\ee
Applying the averaging procedure (A6) to equation (A15), we get
\be
\tilde\rho {d\left<\hat\ell\right>\over dt}
=-{1\over r^2\sin\theta}{\partial\over\partial x_\beta} \left<R_{3\beta}\right>
-\left<\tilde\rho{\partial\over\partial\hat\phi}\hat\Phi\right>,
\ee
where we have used $\left<\partial\tilde p/\partial\phi\right>=0$.
In general,
\be
\left<R_{\alpha\beta}\right>=\delta_{\alpha\beta}\left<J\hat r^2\sin\hat\theta\hat p-r^2\sin\theta\tilde p\right>
-\left<\hat r^2\sin\hat\theta\hat p{\partial\xi_\gamma\over\partial x_\alpha}K_{\gamma\beta}\right>
\ee
is called the radiation stress tensor (e.g., Grimshaw 1984).

Equation (A18) may be regarded as the $\phi$ component of the meanflow equation, the left-hand-side of which may represent
the time evolution of the meanflow and the right-hand-side the forcing by the waves represented by $\xi_\alpha$.
So far we have not assumed that the amplitudes of the waves $\xi_\alpha$ are infinitesimally small,
and in principle we can formulate the wave-meanflow interaction as a nonlinear theory, which includes 
equations of motion for both the meanflows and waves.
To avoid solving such a difficult non-linear problem,
we use a linear theory to describe waves $\xi_\alpha$ and we are satisfied with calculating the forcing terms in
the meanflow equation using the linear waves $\xi_\alpha$.

If we employ a linear theory to describe waves represented by the displacement $\xi_\alpha$, 
we may write
\be
\hat p=\bar p+\delta p=\bar p+p'+{\xi_\alpha}{\partial \bar p\over\partial x_\alpha},
\ee
where $\bar p$ is the pressure in equilibrium state, $\delta p$ and $p'$ denote the Lagrangian and Eulerian
perturbations, respectively.
Applying the averaging procedure (A6), to second order of the perturbations
we obtain after some manipulations for the second term on the right hand side of equation (A16)
\be
 \left\langle {\hat r^2 \sin \hat\theta \hat p\frac{{\partial \xi _\gamma  }}{{\partial x_3 }}K_{\gamma \alpha } } \right\rangle   
  = \left\langle {r^2 \sin \theta \frac{{\partial \xi _\alpha  }}{{\partial x_3 }}p'} \right\rangle  + \left\langle {\frac{\partial }{{\partial x_\gamma  }}\left( {r^2 \sin \theta \frac{{\partial \xi _\alpha  }}{{\partial x_3 }}\xi _\gamma  \bar p} \right)}\right\rangle ,
\ee
where we have used
$\left<{\partial f/\partial\phi}\right>=\partial\left<f\right>/\partial\phi=0$
and $\partial\bar p/\partial\phi=0$ for the equilibrium pressure $\bar p$, and $x_3=\phi$.
Because
\begin{eqnarray}
\displaystyle
{\partial\over\partial x_\alpha}{\partial\over\partial x_\gamma}\left(\bar p r^2\sin\theta{\partial\xi_\alpha\over\partial \phi}
\xi_\gamma\right)
\displaystyle
&=&{\partial\over\partial\phi}\left[{\partial\over\partial x_\alpha}{\partial\over\partial x_\gamma}
\left(\bar p r^2\sin\theta\xi_\alpha\xi_\gamma\right)\right]\nonumber\\
&-&{\partial\over\partial x_\alpha}{\partial\over\partial x_\gamma}
\left(\bar p r^2\sin\theta\xi_\alpha{\partial\xi_\gamma\over\partial\phi}\right),
\end{eqnarray}
we find
\be
\left<{\partial\over\partial x_\alpha}{\partial\over\partial x_\gamma}\left(\bar p r^2\sin\theta{\partial\xi_\alpha\over\partial \phi}\xi_\gamma\right)\right>
=\left<{\partial\over\partial x_\alpha}{\partial\over\partial x_\gamma}
\left(\bar p r^2\sin\theta\xi_\alpha{\partial\xi_\gamma\over\partial\phi}\right)\right>=0,
\ee
and hence we can omit the second term on the right-hand-side of equation (A21) to obtain
\be
 \left\langle {\hat r^2 \sin \hat\theta \hat p\frac{{\partial \xi _\gamma  }}{{\partial x_3 }}K_{\gamma \alpha } } \right\rangle  =    \left\langle {r^2 \sin \theta \frac{{\partial \xi _\alpha  }}{{\partial x_3 }}p'} \right\rangle ,
\ee
and hence
\be
\left<R_{3\alpha}\right>=\delta_{3\alpha}\left<J\hat r^2\sin\hat\theta\hat p-r^2\sin\theta\tilde p\right>
-\left\langle {r^2 \sin \theta \frac{{\partial \xi _\alpha  }}{{\partial x_3 }}p'} \right\rangle .
\ee
The meanflow equation (A18) is then reduced to
\be
\bar\rho{d\left<\hat\ell\right>\over dt}=-\nabla\cdot\left<\pmb{\xi}{\partial p^\prime\over\partial\phi}\right>
-\left<\bar\rho{\partial\hat\Phi\over\partial\hat\phi}\right>,
\ee
where $\pmb{\xi}=\xi_r\pmb{e}_r+\xi_\theta\pmb{e}_\theta+\xi_\phi\pmb{e}_\phi$ with $\pmb{e}_r$, $\pmb{e}_\theta$,
and $\pmb{e}_\phi$ being the orthonormal vectors in the $r$, $\theta$, and $\phi$ directions, and
$\xi_r=\xi_1$, $\xi_\theta=r\xi_2$, and $\xi_\phi=r\sin\theta\xi_3$,
and $\tilde \rho=\bar \rho$.
Since
\be
\left<\bar\rho{\partial\hat\Phi\over\partial\hat\phi}\right>=
\left<\bar\rho{\partial x_\beta\over\partial\hat\phi}{\partial\over\partial x_\beta}\left(\bar\Phi+\Phi'+\pmb{\xi}\cdot\nabla\bar\Phi\right)\right>=\nabla\cdot\left<\bar\rho\pmb{\xi}{\partial\Phi'\over\partial\phi}\right>
+\left<\rho'{\partial\Phi'\over\partial\phi}\right>,
\ee
which is correct to second order of perturbations, we obtain
\be
\bar\rho{d\left<\hat\ell\right>\over dt}=-\nabla\cdot\left<\pmb{\xi}{\partial p^\prime\over\partial\phi}
+\bar\rho\pmb{\xi}{\partial\Phi^\prime\over\partial\phi}+{\nabla\Phi^\prime\over 4\pi G}{\partial\Phi^\prime\over\partial\phi}\right>,
\ee
where we have used $\partial\bar\Phi/\partial\phi=0$, $\nabla^2\Phi^\prime=4\pi G\rho^\prime$ and
$\left<\nabla\Phi^\prime\cdot\partial\nabla\Phi^\prime/\partial\phi\right>=0$.
Since the azimuthal and temporal dependence of the perturbations are assumed to be given by the factor $\exp\left({\rm i}m\phi+{\rm i}\omega t\right)$ with $m$ and $\omega$ being the azimuthal wavenumber and oscillation frequency, using, for example,
\be
\left<\xi_\phi\xi_r\right>={1\over 2}{\rm Re}\left(\xi_\phi^*\xi_r\right)={1\over 2}{\rm Re}\left(\xi_\phi\xi_r^*\right),
\ee
we can rewrite equation (A28) as
\begin{eqnarray}
\bar\rho{d\left<\hat\ell\right>\over dt}&=&-{1\over 2}{\rm Re}\left[\nabla\cdot\left(\pmb{\xi}^*{\partial p^\prime\over\partial\phi}
+\bar\rho\pmb{\xi}^*{\partial\Phi^\prime\over\partial\phi}+{\nabla\Phi^{\prime *}\over 4\pi G}{\partial\Phi^\prime\over\partial\phi}\right)\right] \nonumber\\
&=&{m\over 2}{\rm Im}\left[\nabla\cdot\left(\pmb{\xi}^*p^\prime+\bar\rho\pmb{\xi}^*\Phi^\prime
+{\nabla\Phi^{\prime *}\over 4\pi G}\Phi^\prime\right)\right],
\end{eqnarray}
where the asterisk indicates the complex conjugation.

Integrating over a spherical surface, we obtain
\be
\bar\rho\overline{{d\left<\hat\ell\right>\over dt}}={m\over 2 r^2}{d\over d r}r^2
{\rm Im}\left(\overline{\xi_r^*p^\prime+\bar\rho\xi_r^*\Phi^\prime+{\partial\Phi^{\prime *}\over \partial r}{\Phi^\prime\over 4\pi G}}\right),
\ee
where $\overline f=(4\pi)^{-1}\int_0^{2\pi}\int_0^\pi f\sin\theta d\theta d\phi$.
This expression (A31) is essentially the same as that used by Papaloizou \& Savonije (1997) who discussed 
the forcing by low frequency modes tidally excited in a massive star by the orbital motion of the companion
in a binary system
(see also Ryu \& Goodman 1992, Lin, Papaloizou, \& Kley 1993).
In the Cowling approximation (Cowling 1941), we obtain
\be
\bar\rho\overline{{d\left<\hat\ell\right>\over dt}}={m\over 2\pi r^2}{d W(r)\over d r},
\ee
where  
\be
W(r)=-\pi r^2{\rm Im}\left(\overline{p^{\prime *}\xi_r}\right)
\ee
may be regarded as a work function (e.g., Unno et al. 1989).
We note that $dW/dr>0$ and $dW/dr<0$ respectively indicate the
excitation and damping regions for the oscillation modes.
For uniformly rotating stars, 
non-axisymmetric ($m\not=0$) oscillations of rotating stars
are separated into prograde and retrograde modes, and
in our convention, positive (negative) $m$ is used for retrograde (prograde) modes.
From equation (A32), we find that 
there occurs acceleration (deceleration) of the zonal flow in
the damping (excitation) regions of prograde modes ($m<0$),
while deceleration (acceleration) occurs in the damping (excitation) regions of retrograde modes ($m>0$).

It is possible to rewrite the mean flow equation (A26) using the Reynolds stress.
Substituting into equation (A26) the $\phi$ component of the linearized momentum equation given by
\be
{\partial p^\prime\over\partial\phi}=-\bar\rho r\sin\theta
\left[\left({\partial\over\partial t}+{\bar v_\phi\over r\sin\theta}{\partial\over\partial\phi}\right)v_\phi^\prime
+{1\over r\sin\theta}{\partial \bar \ell\over\partial r}v_r^\prime
+{1\over r\sin\theta}{1\over r}{\partial \bar \ell\over\partial\theta}v_\theta^\prime
+{1\over r\sin\theta}{\partial\Phi^\prime\over\partial\phi}\right],
\ee
where 
$
\bar v_\phi=r\sin\theta\Omega_1(r,\theta),
$
$
\bar \ell=\Omega r^2\sin^2\theta,
$
$
\Omega=\Omega_1+\Omega_c,
$
and using
\be
\pmb{v}^\prime={\partial\pmb{\xi}\over\partial t}+\Omega_1\sum_j{\partial\xi_j\over\partial\phi}\pmb{e}_j
-r\sin\theta\left(\pmb{\xi}\cdot\nabla\Omega_1\right)\pmb{e}_\phi,
\ee
we obtain after some manipulations
\begin{eqnarray}
\bar\rho{d\left<\hat\ell\right>\over dt}&=&-{1\over r^2}{\partial\over\partial r}
\left<\bar\rho r^3\sin\theta
 v_r^{\prime }\left( v_\phi^\prime  + 2 \Omega f_\theta\cos\theta \xi _\theta 
    \right)\right>\nonumber\\
&-&{1\over r\sin\theta}{\partial\over\partial\theta}
\left<\bar\rho r\sin^2\theta v_\theta^{\prime }\left( v_\phi^\prime 
 + 2\Omega f_r\sin \theta \xi _r   \right)\right>
 -\left<\rho'{\partial\Phi'\over\partial\phi}\right>
+{\partial\over\partial t}\nabla\cdot\left<\bar\rho r\sin\theta Q\pmb{\xi}\right> \nonumber \\
&-&{\partial\over\partial t}\left({1\over r^2}{\partial\over\partial r}\left<\bar\rho r^3\sin^2\theta\Omega f_r\xi_r^2\right>
+{1\over r\sin\theta}{\partial\over\partial\theta}\left<\bar\rho r\sin^2\theta\cos\theta\Omega f_\theta\xi_\theta^2\right>\right),
\end{eqnarray}
where 
\be
f_r=1+{1\over 2}{\partial\ln\Omega\over\partial\ln r}, \quad f_\theta=1+{1\over 2}{\partial\ln\Omega\over\partial\ln\sin\theta},
\ee
\be
Q=v_\phi^\prime+2\Omega f_\theta\cos\theta\xi_\theta+2\Omega f_r\sin\theta\xi_r.
\ee
If we neglect the terms like $\partial\left<Q\pmb{\xi}\right>/\partial t$ and
$\partial\left<\xi_r^2\right>/\partial t$ assuming
the amplitudes of the perturbations are saturated, for example, by nonlinear effects,
the expression for the forcing terms in the mean flow equation (A36) with $f_\theta=1$ reduce to that proposed by 
Pantillon et al. (2007) and Mathis (2009).

It is instructive to give an example of the forcing term in the meanflow equation.
To see the behavior of the forcing term, we introduce the local timescale $\tau^{\rm AM}$ defined by
\be
{1\over\tau^{\rm AM}}\equiv{\overline{d\left<\ell\right>/dt}\over\overline{\left<\ell\right>}}.
\ee
An example of $1/\tau^{\rm AM}$ is given for low frequency modes of
a $4M_\odot$ main sequence star with $X=0.7$ and $Z=0.02$ in Figure 6, where the model has been calculated 
with a standard stellar evolution code with the OPAL opacity (Iglesias \& Rogers 1996), and we have 
used the method of calculation given by Lee \& Saio (1993) for non-adiabatic oscillation modes of
a uniformly rotating star.
As the figure shows, there occurs a strong acceleration by retrograde $g$- and $r$-modes
in the layers at $r/R_*\sim 0.95$, although the prograde $g$-mode contributes to deceleration of the
rotation.

\begin{figure}
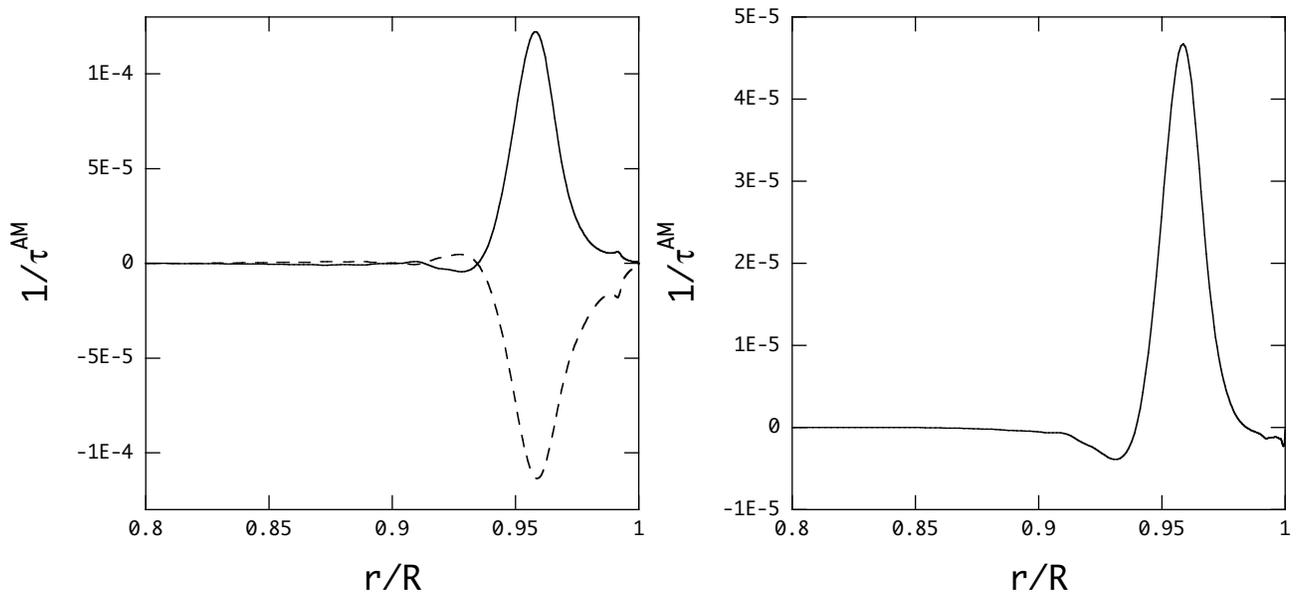

\resizebox{0.5\columnwidth}{!}{
\includegraphics{f6a.epsi}}
\resizebox{0.5\columnwidth}{!}{
\includegraphics{f6b.epsi}}
\caption{$1/\tau^{\rm AM}$ for unstable $l=|m|=1$ $g_{22}$ modes for $\Omega_c/(GM_*/R_*^3)^{1/2}=0.1$ (left panel)
and an unstable $l=|m|+1=2$ $r_{30}$ mode for $\Omega_c/(GM_*/R_*^3)^{1/2}=0.4$ (right panel)
of a $4M_\odot$ main sequence model with $X=0.7$, $Z=0.02$, and  the central hydrogen content $X_c=0.3646$, where
$M_*$ and $R_*$ denote the mass and radius of the model.
The solid and dotted lines indicate retrograde and prograde modes, respectively,
and $\tau^{\rm AM}$ is given in seconds.
The amplitude normalization is given by setting the radial displacement associated with the spherical harmonic function
$Y_{l=|m|}^m(\theta,\phi)$ equal to $R_*$ at the surface.
Note that uniform rotation is assumed.
}
\end{figure}

\end{appendix}

\end{document}